\documentclass[aps,showpacs,twocolumn,prl]{revtex4}
\usepackage{amssymb}
\usepackage{graphicx}
\usepackage{amsmath}
\begin{document}
\title{ Thermalization of random motion in weakly confining potentials}
\author{Piotr Garbaczewski and   Vladimir Stephanovich}
\affiliation{Institute of Physics, University of Opole, 45-052 Opole, Poland}
\date{\today}
\begin{abstract}
We show that in  weakly confining conservative force fields, a subclass of diffusion-type (Smoluchowski) processes,
admits a family of "heavy-tailed" non-Gaussian equilibrium probability density functions (pdfs),
with none or a finite number of moments. These  pdfs, in the standard Gibbs-Boltzmann form,  can be  also
 inferred  directly  from an extremum  principle,   set  for Shannon entropy under a constraint   that the mean
  value of the  force potential  has been a priori prescribed.
 That enforces  the  corresponding Lagrange multiplier  to  play  the  role of   inverse temperature.
Weak confining properties of the  potentials are  manifested in a thermodynamical peculiarity  that  thermal equilibria
can be approached  \it   only  \rm in a bounded temperature interval $0\leq T < T_{max} =2\epsilon _0/k_B$, where $\epsilon _0$
 sets an energy scale. For $T \geq  T_{max}$ no  equilibrium  pdf  exists.
\end{abstract}
\maketitle

We depart from a folk lore assumption that  a thermodynamical system has  a state characterized   by a  probability density, \cite{mackey0}.
 That amounts to studying (random)  dynamical systems in terms of time-dependent probability density functions (pdfs) and
  discovering whether and   how  the system may approach a state of thermodynamical equilibrium. Localization properties  of pdfs,
   both far from and at equilibrium,  can be  quantified in terms of Shannon entropy. Admissible dynamical equilibria can be inferred
   from  various variational principles.

At this point we invoke  a  classification of maximum entropy principles (MEP) as given in Ref.~\cite{kapur}. Let us
look for pdfs that derive from so-called first inverse MEP:  given a  pdf $\rho (x)$, choose
an \it appropriate \rm set of constraints such that  $\rho (x)$ is  obtained if Shannon  measure of  entropy  is maximized
(strictly speaking, extremized) subject to those constraints.

Namely if a system evolves in a potential
$V(x)$ (at the moment,  we consider a coordinate $x$ to be dimensionless), we can introduce the following functional
\begin{equation} \label{fx1}
L \{ \rho (x) \}=-\lambda \int_{-\infty}^\infty V(x) \rho (x) dx -  \int_{-\infty}^\infty \rho (x) \ln [\rho (x)]dx,
\end{equation}
where (without limitation of generality) we consider the  one-dimensional case. The first term comprises  the mean value of a potential.
 A constant $\lambda$ is (as yet  physically unidentified) Lagrange multiplier, which takes care of aforementioned constraints. The  second term   stands for  Shannon entropy of a continuous (dimensionless) pdf $\rho (x)$.
  An extremum of the functional $L \{ \rho (x) \}$  can be found by means of standard variational arguments and gives rise to  the
   following  general form of an   extremizing  pdf $\rho _*(x)$
\begin{equation} \label{fx2}
\rho _*(x)=C\exp(-\lambda V(x)),
\end{equation}
which, if regarded as the Gibbs-Boltzmann pdf,   implies that the  parameter $\lambda$ can be interpreted  as inverse temperature,
$\lambda=(k_BT)^{-1}$ ($k_B$ is Boltzmann constant),  at which a state of equilibrium (asymptotic pdf) is reached by a
  random dynamical  system in a confining potential   $V(x)$.

A deceivingly simple question has been posed in chap. 8.2.4 of Ref. \cite{kapur}.   Having dimensionless logarithmic
potential  ${\cal{V}}(x)= \ln (1+x^2)$, one should  begin with evaluating a mean value
 ${\cal{U}}= \langle {\cal{V}}\rangle$  $\equiv \int_{-\infty}^\infty {\cal{V}}(x) \rho (x) dx$.  Next one needs to  show that
only if \it  this particular value \rm  is prescribed in the above MEP procedure,  Cauchy distribution will ultimately arise.
Additionally,  one should  answer  what kind of distribution would arise if \it  any other  \rm  positive expectation value is chosen.
The answer proves not  to be that  straightforward and we shall analyze this issue below.

To handle the problem  we admit all  pdfs $\rho (x)$ for which the mean value $\langle \ln (1+x^2) \rangle $ exists,
 i.e. takes  whatever finite positive value.  Then,  we adopt  the previous  variational procedure with the use of
Lagrange multipliers.  This procedure shows that what we extremize is not the (Shannon) entropy itself,
but a functional  ${\cal{F}}$ with a clear thermodynamic connotation  (Helmholtz free energy analog, \cite{stat}):
 \begin{equation} \label{fe1}
\Phi  (x) = \alpha \,   {\cal{V}}(x) + \ln \rho (x)  \rightarrow
{\cal{F}} = \langle \Phi \rangle  = \alpha \, \langle
{\cal{V}}\rangle - {\cal{S}}(\rho )\, .
\end{equation}
Here ${\cal{S}}(\rho )= - \langle \ln \rho \rangle $ and $\alpha $  is a Lagrange
multiplier. From now on we,  we consider
a  parameter multiplier in the form  $\alpha=\epsilon_0/(k_BT)$, where $\epsilon_0$
is characteristic energy scale of a system.  Note that Eq.~(3), provides a dimensionless
 version of   a familiar formula  $F=U-TS$,   relating  the Helmholtz free energy $F$,
internal energy $U$ and entropy $S$ of a random dynamical system.

The extremum condition for ${\cal{F}}$
\begin{equation} \label{ecf}
{\frac{\delta {\cal{F}}(\rho )}{\delta \rho }}=0
\end{equation}
yields an extremizing pdf in the form
\begin{equation}\label{ecf1}
\rho _{\alpha  }(x) = {\frac{1}{Z_{\alpha }}}\, (1+x^2)^{-\alpha }
\end{equation}
provided the normalization factor  $Z_{\alpha } = \int_{-\infty}^\infty (1+x^2)^{-\alpha
}\, dx $ exists.  It turns out that the integral can be evaluated explicitly in terms of
$\Gamma$ - functions
\begin{equation}\label{za}
Z_{\alpha } =\frac{\sqrt{\pi}\Gamma(\alpha - 1/2)}{\Gamma(\alpha)},\ \alpha>1/2 \, ,
\end{equation}
so that we can in principle deduce  a
numerical  value of the  parameter $\alpha$,  by resorting
to our  assumption that the mean value $\langle {\cal{V}} \rangle
_{\alpha }$ has  actually  been fixed.

\begin{figure}
\begin{center}
\includegraphics [width=0.9\columnwidth]{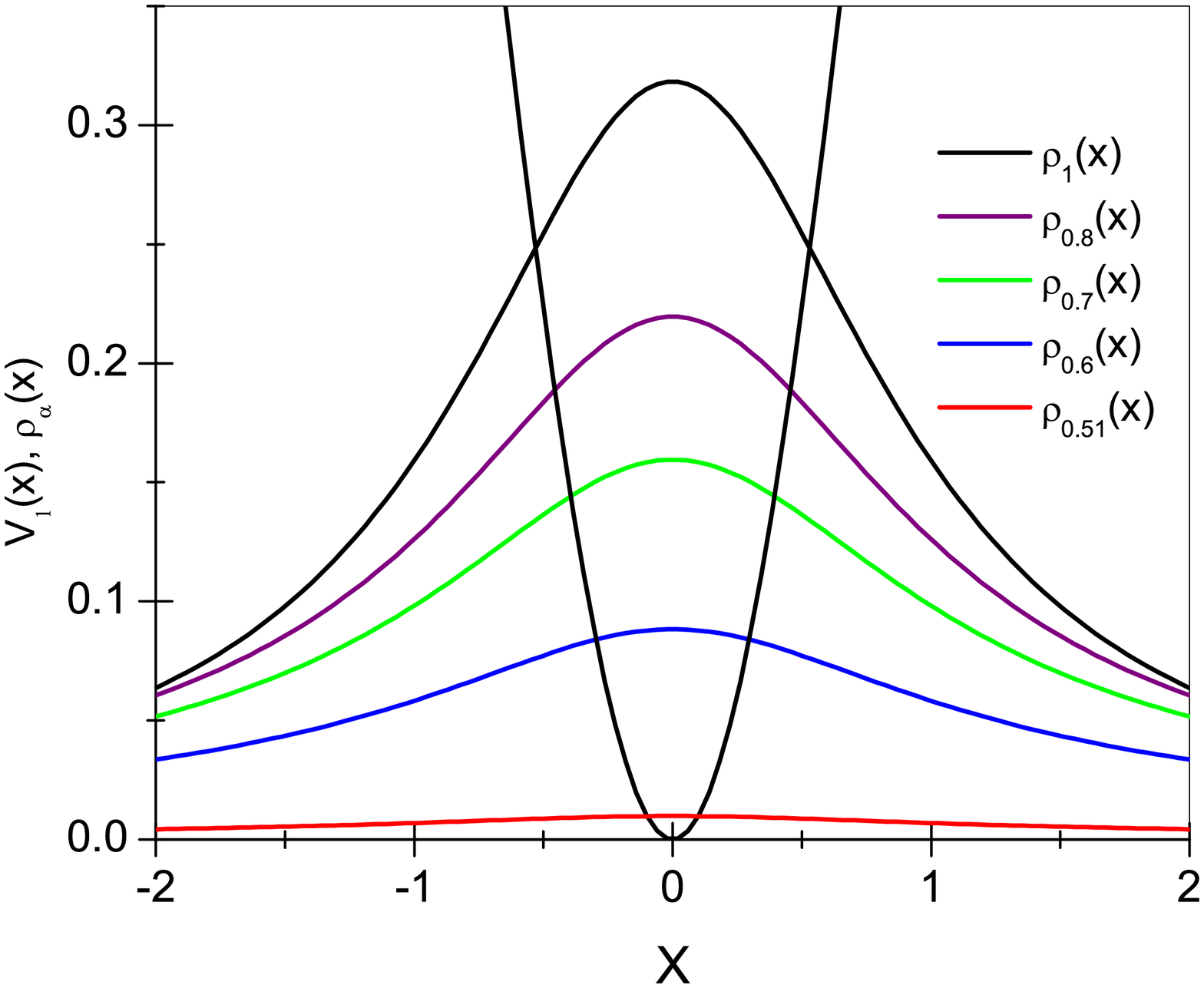}
\end{center}
\caption{$\ln (1+x^2)$   against $\rho _{\alpha }(x)$ with $1/2<\alpha \leq 1$.}\label{fig:sew}
\end{figure}

\begin{figure}
\begin{center}
\includegraphics [width=0.9\columnwidth]{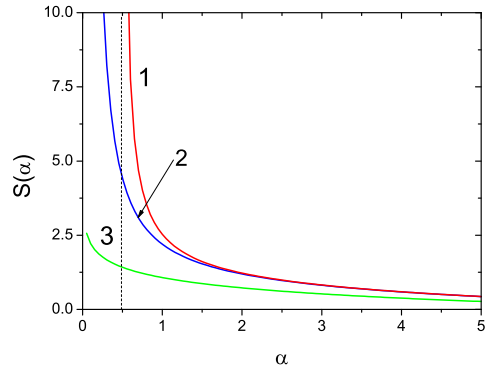}
\end{center}
\caption{${\cal{U}}_{\alpha }$  (curve 1) and its asymptotic expansion at large $\alpha$  (curve 2).}\label{fig:sev}
\end{figure}

We observe that the extremizing  pdfs  $\rho _{\alpha }$  \eqref{ecf1} appear  to  constitute a one-parameter
family of pdfs, named by us Cauchy family, \cite{gar,stef}. It is seen from Eq. \eqref{za} that the extremizing
functions \eqref{ecf1} are normalizable only for $\alpha >1/2$ as at $\alpha \to1/2$ we have  $Z_\alpha \to \infty$.
We illustrate this behavior in Fig.~\ref{fig:sew}, which shows that as $\alpha \to1/2$ the whole pdf $\rho_\alpha(x)$ becomes
 identically equal  to  zero.
To deduce the values of above dimensionless inverse temperature $\alpha$, we need an explicit expression for the  mean value
\begin{equation}\label{ln1}
 {\cal{U}}_{\alpha }=  \langle {\cal{V}}\rangle _{\alpha }  =
\frac{\Gamma(\alpha)}{\sqrt{\pi}\Gamma(\alpha - 1/2)}\int_{-\infty}^{\infty}\frac{\ln(1+x^2)}{(1+x^2)^\alpha}dx.
\end{equation}
It turns out that it can  be given  in terms of digamma function $\psi(x)=d(\ln \Gamma)/dx$ \cite{abr}. We  get
\begin{equation}
{\cal{U}}_{\alpha}=
-\frac{2\pi}{\sin(2\pi\alpha)}+\psi(1-\alpha)-\psi\left(\frac 32-\alpha\right),\ \alpha>\frac 12.\label{ln3}
\end{equation}
The dependence of  ${\cal{U}}_{\alpha}$  on $\alpha $ is reported in Fig.~\ref{fig:sev}, where it is seen that this function is divergent at
$\alpha =1/2$ (see also below) and decays monotonously at large $\alpha$.
This decay  is conrolled by an asymptotic expansion
\begin{equation}\label{as}
{\cal{U}}_{\alpha}  \approx   \frac{1}{2\alpha}+\frac{3}{8\alpha^2}+\frac{1}{4\alpha^3}+...,
\end{equation}
which is also shown on Fig. \ref{fig:sev}. It is seen that decay of ${\cal{U}}_{\alpha}$ at large $\alpha$  obeys the inverse power law. It follows from Fig. \ref{fig:sev} that the expansion \eqref{as} gives
a very good approximation of ${\cal{U}}_{\alpha}$ for $\alpha >3$.

We note that, apparently,  Eq. \eqref{ln3}  involves another  divergence problems,  if we choose   integer  $\alpha $.
This obstacle can be circumvented by transforming Eq. \eqref{ln3} to an equivalent form that has no
 (effectively  removable) divergencies. Namely,  we get
\begin{equation} \label{gl3}
{\cal{U}}_{\alpha}=-\pi \tan \pi \alpha +\psi(\alpha)-\psi\left(\frac 32-\alpha\right)
\end{equation}
and the  tangent  contribution vanishes for integer $\alpha $.
 On the other hand,    this expression shows that the divergence of ${\cal{U}}_{\alpha}$ at $\alpha \to 1/2$  originates
  from the first term in \eqref{gl3},  as $\psi$ functions have finite values at
this point. Near $\alpha =1/2$ the first term of Eq. \eqref{gl3} diverges as $(\alpha-1/2)^{-1}$,
 which is clearly seen in  Fig. \ref{fig:sev}.

The preceding discussion shows that weak confinement properties of logarithmic potential manifest themselves in the
fact that the corresponding (Cauchy family) equilibrium  pdfs exist only for the semi-infinite range of (dimensionless)
 inverse temperatures $\alpha \in (1/2,+\infty )$.
  Note that for strongly confining potentials (like e.g.  $V(x) \sim x^2$ or $x^4$) this is not the case as it follows
  from Eq. \eqref{fx2} that  the normalizing integral $C^{-1}=\int_{-\infty}^\infty \exp[-\lambda V(x)]dx$ is convergent
  at any $\lambda$,  so that  the whole temperature range $[0,\infty )$ is allowed.

Let us turn to a quantitative   discussion of an impact of weak confinement properties of the  above logarithmic potentials
   on thermodynamical properties of a random system.
To this end, we should focus our attention on
 the thermodynamic meaning of the (originally Lagrange multiplier)  parameter $\alpha$  and its  semi-bounded range of
 variability $(1/2,+\infty )$.  This is also related to above problem of explicit calculation of $\alpha$ for a specific pdf.

\begin{figure}
\begin{center}
\includegraphics [width=0.9\columnwidth]{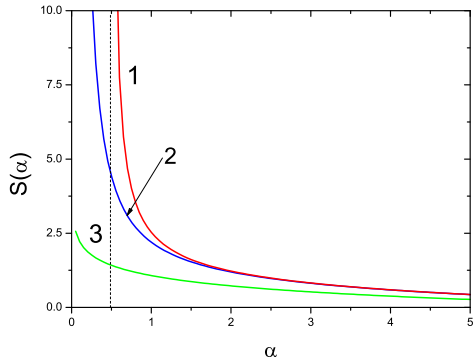}
\end{center}
\caption{ ${\cal{S}}_{\alpha }$  for Cauchy family (curve 1) and its asymptotic expansion at large $\alpha$  (curve 2) .
${\cal{S}}^G_{\alpha }$ for Gaussian family  is also shown (curve 3).}\label{fig:sei}
\end{figure}
With an explicit expression for   Cauchy family pdfs \eqref{ecf1} in hands,  we readily  evaluate  Shannon entropy to obtain
\begin{equation}
{\cal{S}}_{\alpha }=-\int_{-\infty}^{\infty}\rho_\alpha (x)\ln \rho_\alpha (x)dx=
  \ln Z_\alpha+\alpha \,  {\cal{U}}_{\alpha}. \label{sa}
\end{equation}
Then, the  (as yet dimensionless) Helmholtz free energy ${\cal F}_{\alpha }$   reads
\begin{equation} \label{fedim}
{\cal {F}}_{\alpha }=   \alpha {\cal{U}}_{\alpha }  -  {\cal{S}}_{\alpha } \equiv -\ln Z_{\alpha } ,
\end{equation}
with $\alpha $  being the dimensionless inverse temperature, c.f.  Ref. \cite{stat}.

We note, that in view of the divergence of $Z_{\alpha }$, both the Shannon entropy and the Helmholtz  free energy
 (likewise ${\cal{U}}_{\alpha }$) cease to exist
at $\alpha = 1/2$.
We plot  ${\cal{S}}_{\alpha }$   as a function of $\alpha $ in Fig.\ref{fig:sei}. It is seen that entropy  monotonously
 decays for $\alpha >1/2$ and for larger values
of $\alpha $. An   asymptotic expansion of the entropy   shows  logarithmic plus inverse power signatures
\begin{equation}\label{sas}
{\cal{S}}_{\alpha } \approx \frac 12\left(1-\ln \frac{\alpha}{\pi}\right) +\frac{3}{4\alpha}+\frac{3}{8\alpha^2}+...
\end{equation}
These series are shown along with the  entropy in Fig.~\ref{fig:sei}.

As $\alpha $   grows, the number of moments of respective pdfs increases. That allows to expect that  an "almost Gaussian" behavior should be displayed by $\alpha \gg 1$ members of Cauchy family \eqref{ecf1}. This is indeed the case. To this end, let us recall that the normalized Gaussian function with  zero mean and  variance   $\sigma ^2 =1/2\alpha$  reads
\begin{equation}\label{ga1}
    \rho_{\alpha }^G(x)=\sqrt{\frac{\alpha}{\pi}}\ e^{-\alpha x^2}\equiv {\frac{1}{Z_{\alpha }^G}}  e^{-\alpha x^2}.
\end{equation}
The corresponding Shannon entropy gives exactly the first term of Eq. \eqref{sas}
\begin{equation}
{\cal{S}}^G_{\alpha }=  \frac 12\ln \frac{\pi e}{\alpha}.\label{eng}
\end{equation}
The plot of  ${\cal{S}}^G_{\alpha }$ is reported in Fig.\ref{fig:sei} together with   ${\cal{S}}_{\alpha }$ for   Cauchy family. This equality suggests that at large $\alpha$ Cauchy family pdfs \eqref{ecf1} transit to Gaussian one \eqref{ga1}. This transition can be revealed if we rewrite  Cauchy family pdfs \eqref{ecf1} in the exponential   form
   $ \rho_\alpha (x)=  \exp (\ln \rho_\alpha)$ where $\ln \rho_\alpha=-\ln Z_{\alpha }-\alpha \ln(1+x^2)$.   It can be shown that at large $\alpha$ the "tails" of pdf do not make substantial contribution so that only small $x$ play a role.
This means that in limiting case of large $\alpha$ we can expand $\ln(1+x^2)$ at small $x$.
Then, employing series expansions
\begin{equation}
Z_{\alpha \to \infty}^{-1} \approx \sqrt{\frac{\alpha}{\pi}}-\frac{3}{8\sqrt{\pi \alpha}}-\frac{7}{128\alpha\sqrt{\pi \alpha}}-...\label{lny2}
\end{equation}
we arrive at
\begin{equation}\label{tf2}
    \rho _{\alpha }(x)\approx\left[\sqrt{\frac{\alpha}{\pi}}-\frac{3}{8\sqrt{\pi \alpha}}...\right]\,
\exp\left(-\alpha x^2 +\alpha \frac{x^4}{2}...\right)
\end{equation}
whose leading order  gives  exactly the normalized  Gaussian  \eqref{ga1}.
Accordingly, c.f. Fig.~\ref{fig:sei}, an  inequality $\alpha >5$  sets a regime where a fairly good  approximation of the Cauchy family pdfs by Gaussian is valid.   Note that ultimately,  in the $\alpha \to \infty $ limit, both families  of functions  (sequentially)  approach  the Dirac delta function(al).

\begin{figure*}
\begin{center}
\includegraphics [width=0.66\columnwidth]{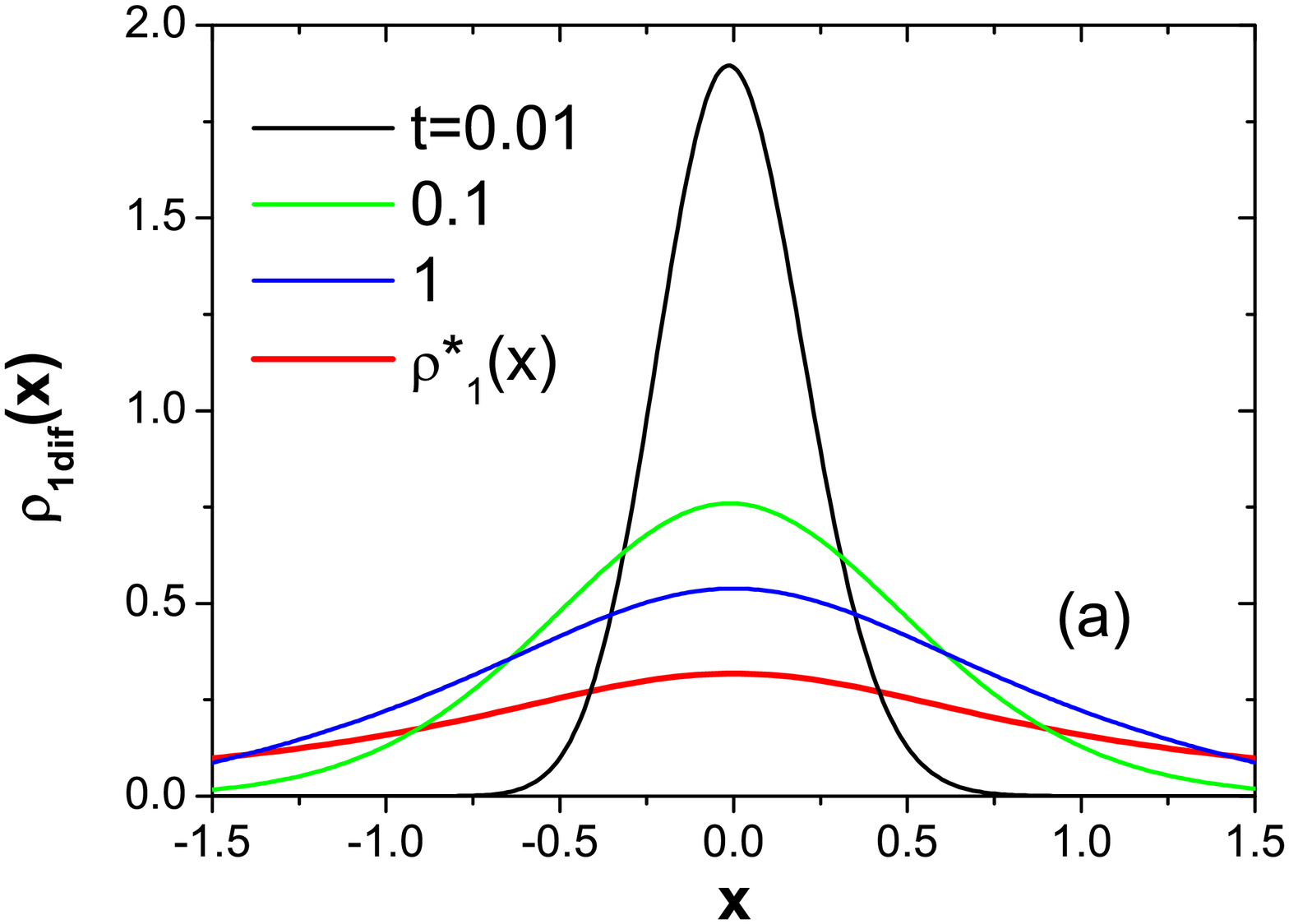}
\includegraphics [width=0.66\columnwidth]{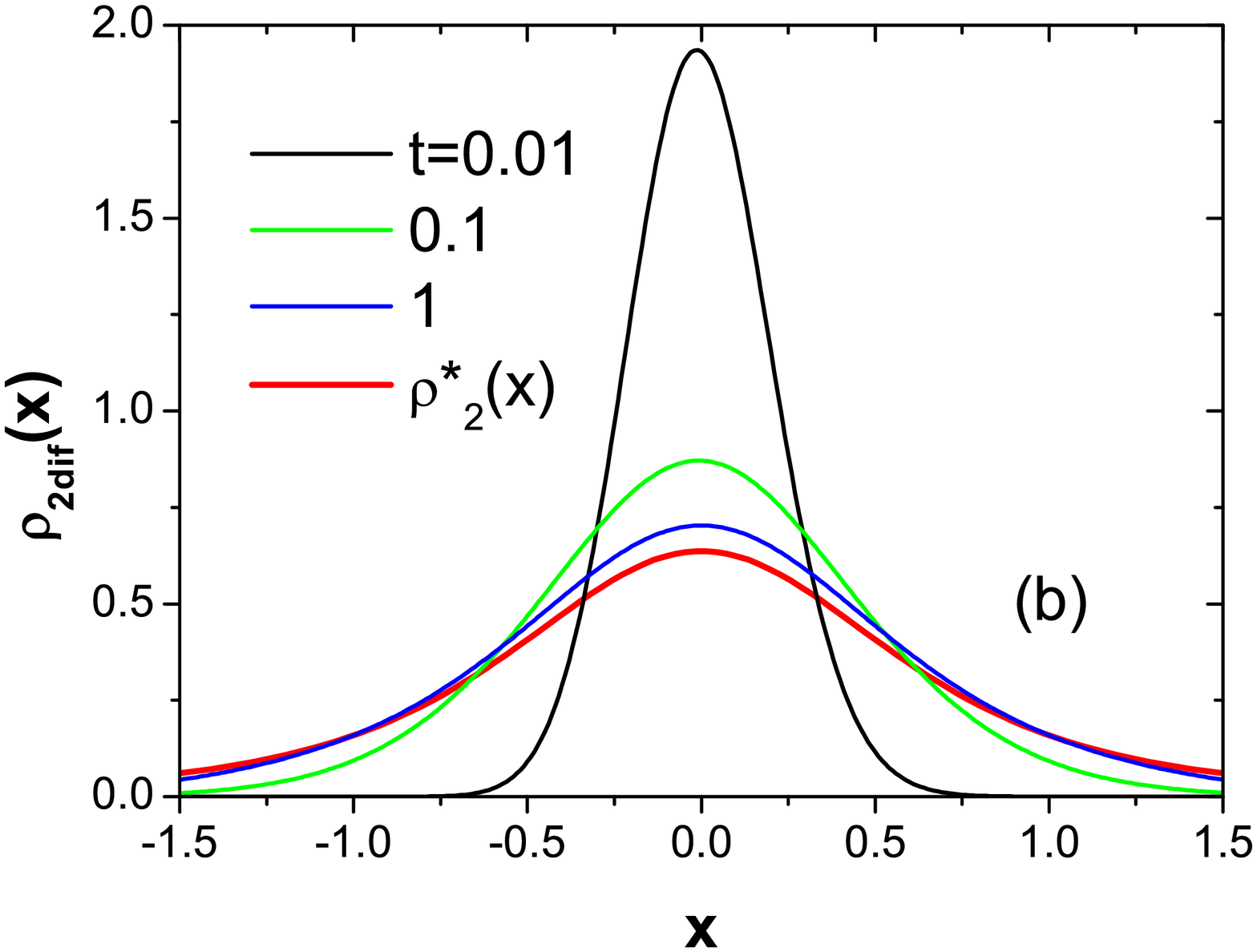}
\includegraphics [width=0.66\columnwidth]{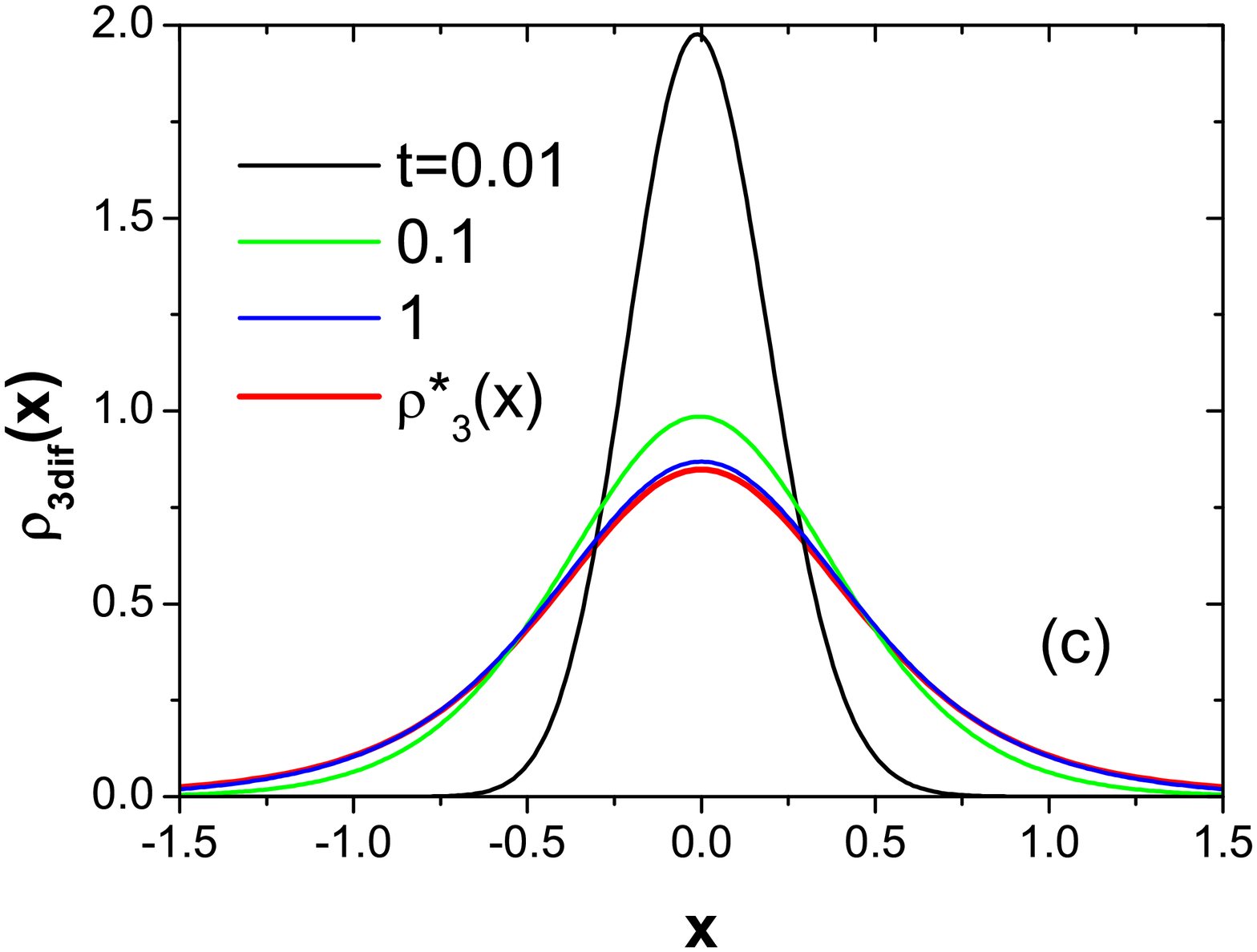}
\end{center}
\caption{Time evolution of pdf's $\rho(x,t)$ for Smoluchowski  processes in logarithmic potential
$\ln(1+x^2)$. The initial ($t=0$) pdf is set
to be  a Gaussian with height 25 and half-width $\sim 10^{-3}$. The first depicted stage of evolution corresponds to  $t=0.01$.  Target pdfs are the members of Cauchy family \eqref{ecf1} for $\alpha=1$ (panel (a)), 2 (panel (b))  and 3 (panel (c)) respectively.}\label{fig:zu}
\end{figure*}

Let us now show that the  variational principle \eqref{fx1} explicitly
identifies an  equilibrium solution of the Fokker-Planck equation for standard Smoluchowski diffusion processes.
 To address our thermalization issue correctly, we now use
dimensional units. The Fokker - Planck equation that drives an initial   probability density $\rho(x,t=0)$ to its final (equilibrium) form $\rho(x,t \to \infty)$  reads
\begin{equation} \label{gpe}
\partial _t\rho = D\Delta \rho -  \nabla \cdot ( b(x) \rho ).
\end{equation}
We   still   refer (although without substantial  limitation of generality)  to the  1D case, so that $ \nabla \equiv \partial /\partial x$.
Here, the drift field $b(x)$ is time-independent and conservative, $b(x)=-\nabla V(x)/(m\gamma)$ ($V(x)$ is a potential, while $m$ is a mass and $\gamma$ is a reciprocal relaxation time of a system).
We keep in mind that $\rho$ and $b\rho $ vanish at spatial infinities or other integration interval borders.

If Einstein fluctuation-dissipation relation $D=k_BT/m\gamma $ holds, the equation \eqref{gpe} can be identically rewritten
 in the form $\partial _t\rho = \nabla [\rho \nabla \Psi ]/(m\gamma)$, where
\begin{equation} \label{tyr0}
\Psi = V + k_BT \ln \rho
\end{equation}
whose mean value is indeed the Helmholtz free  energy  of  random  motion
\begin{equation} \label{tyr1}
F \equiv \left< \Psi \right> = U - T S.
\end{equation}
Here the  (Gibbs)  entropy reads $S =  k_B {\cal{S}}$, while   an internal energy is $ U =
\left< V\right>$. In view of assumed boundary restrictions at spatial infinities, we have $
\dot{F}  =     - (m\gamma ) \left<{v}^2\right>  \leq 0 $.
Hence, $F$ decreases as a function of time  towards its  minimum $F_*$,
or  remains constant.

Let us consider the stationary (large time asymptotic) regime
associated with an invariant density $\rho _{*}$ (c.f.  \cite{mackey} for an extended discussion of that issue).
Then,  $\partial _t\rho = 0$  and we have $\nabla \Psi [\rho_*] =C \rho_*$ ($C$ is arbitrary constant) which yields
\begin{equation}\label{tyr2}
\rho _{*} = {\frac{1}Z} \exp[ - V/k_BT]\, .
 \end{equation}
Therefore, at equilibrium:
\begin{equation} \label{tyr3}
\Psi _{*} = V + k_BT \ln \rho _{*}  \Longrightarrow \langle \Psi _{*} \rangle =
 - k_BT \ln Z  \equiv  F_{*},
 \end{equation}
to be compared  with  Eq.~\eqref{fedim}. Here, the  partition function equals $Z= \int \exp(-V/k_BT) dx$, provided that the integral is convergent.
 Since  $Z= \exp (-F_*/k_BT)$ we have  recast  $\rho _*(x)$  in the familiar Gibbs-Boltzmann form  $\rho _* = \exp[(F_* - V)/k_BT]$.

On physical grounds, $V(x)$  carries dimensions of energy. Therefore to establish a physically justifiable
 thermodynamic picture of Smoluchowski diffusion processes, relaxing to Cauchy family pdfs  in their large time asymptotics,
  we need to assume that  logarithmic potentials  ${\cal{V}}(x)=\ln (1+x^2)$  are dimensionally scaled to the form
   $V(x)= \epsilon _0 {\cal{V}}$ where $\epsilon _0$ is an arbitrary constant with physical dimensions of energy.

By employing  $ (1/\epsilon _0)\, V(x) = \ln (1+x^2)$, we can recast  previous variational arguments (MEP procedure) in
terms of dimensional thermodynamical functions. Namely, in view of $k_BT \Phi (x)= V(x)  +k_BT\, \ln \rho (x)$ we have an
 obvious transformation of Eqs.~\eqref{fe1}, \eqref{ecf1} and \eqref{fedim} into Eqs.~\eqref{tyr1},\eqref{tyr2} and \eqref{tyr3}
  respectively with  $ k_BT \Phi (x)= \Psi (x)$.

 The above dimensional arguments tell us that in weakly confining logarithmic potentials, Cauchy family of pdfs  can be regarded as
  a one-parameter family of  \it equilibrium \rm  pdfs, where the reservoir temperature $T$ enters through the exponent $\alpha $.
   Proceeding in this vein,  we note that  $\epsilon _0$ should be regarded as a characteristic energy (energy scale) of
    the considered  random system.

We observe that   $\alpha \to \infty$  corresponds  to $T \to 0$ i.e. a maximal localization (Dirac delta limit)
 of the corresponding pdf. In parallel   ${\cal{S}}(\alpha \to \infty \equiv T \to 0) \to 0$
 and {\em {an analogy with the  Nernst theorem is established}}.

The opposite limiting case $\alpha \to 1/2$ looks interesting as well. Namely, we have
${\cal{S}}(\alpha \to 1/2)\to\infty$. To grasp  the meaning of this limiting regime, we rewrite $\alpha =1/2$ in the form
$k_BT=2\epsilon _0$.  Accordingly, the  temperature scale, within which our system may at all be set at  thermal equilibrium,
 is  bounded:  $0<k_BT<2 \epsilon_0$.
 For temperatures exceeding the  upper bound  $T_{max} = 2\epsilon _0/k_B$  {\em {no thermal equilibrium}} is possible in
  the presence of  (weakly, i.e. weaker then, e.g., $V(x)\sim x^2$ ) confining  logarithmic potentials
   $V(x)=  \epsilon _0\,  \ln (1+x^2)$. The case of $\alpha = 1$ i.e. $k_BT=\epsilon _0$ corresponds
to Cauchy density.

Let us finally add that in Ref.~\cite{gar},  we have analyzed  diffusion processes in logarithmic potentials, with a focus on
 temporal relaxation patterns of the process towards asymptotic pdfs from the Cauchy family. For clarity of presentation
  (dynamical interpolation scenarios between initial Gaussian and equilibrium Cauchy-type pdfs
do not seem to have ever been considered in the literature), in Fig. \ref{fig:zu} we plot various stages of the
diffusive Fokker - Planck dynamics for processes that  all have been  started from a narrow Gaussian.
 It turns out that the   the resultant (large time asymptotic)  equilibrium  pdfs  are members of Cauchy family \eqref{ecf1},
   labeled  respectively by  $\alpha=1$, $2$ and $3$.

\end{document}